\documentclass[5p,times,authoryear]{elsarticle}

\usepackage{ecrc_RIAI}

\usepackage{booktabs}
\usepackage{array}
\usepackage[font = normalsize]{caption}

\usepackage{bm}

\usepackage{multirow}
\usepackage[english]{babel}
\addto\captionsspanish{%
}
\usepackage[latin1]{inputenc}   
\usepackage{amsmath}            
\usepackage{epstopdf}           
\usepackage{flushend}           
\usepackage{setspace}
\usepackage{subfigure}

\usepackage{caption}
\DeclareMathOperator{\Tr}{Tr}

\volume{00}

\firstpage{1}

\runauth{J. Guo et al.}

\usepackage{amssymb}

\usepackage[figuresright]{rotating}
\usepackage{float}

\usepackage{subfigure} 
\usepackage{setspace}

\begin{document}

\captionsetup[figure]{name={Fig.},labelsep=period}
\begin{frontmatter}




\title{Deep neural network based i-vector mapping for speaker verification using short utterances}


 \author[First]{Jinxi Guo\corref{cor1}}
 \ead{lennyguo@g.ucla.edu}

 \author[Second]{Ning Xu}
 \ead{ning.xu@snap.com}

 \author[First]{Kailun Qian}
 \ead{kailunqian@ucla.edu}

 \author[First]{Yang Shi}
 \ead{yangshi5@g.ucla.edu}
 
 \author[First]{Kaiyuan Xu}
 \ead{kyxu@g.ucla.edu}
 
 \author[Third]{Yingnian Wu}
 \ead{ywu@stat.ucla.edu}
 
 \author[First]{Abeer Alwan}
 \ead{alwan@ee.ucla.edu}

 \cortext[cor1]{Corresponding author}

\address[First]{Department of Electrical and Computer Engineering, University of California, Los Angeles, CA, United States, 90095. }
\address[Second]{Snap Inc, Venice, CA, United States, 90063.}
\address[Third]{Department of Statistics, University of California, Los Angeles, CA, United States, 90095.}

\begin{abstract}
Text-independent speaker recognition using short utterances is a highly challenging task due to the large variation and content mismatch between short utterances. I-vector and probabilistic linear discriminant analysis (PLDA) based systems have become the standard in speaker verification applications, but they are less effective with short utterances. In this paper, we first compare two state-of-the-art universal background model (UBM) training methods for i-vector modeling using full-length and short utterance evaluation tasks. The two methods are Gaussian mixture model (GMM) based (denoted I-vector\_GMM) and deep neural network (DNN) based (denoted as I-vector\_DNN) methods. The results indicate that the I-vector\_DNN system outperforms the  I-vector\_GMM system under various durations (from full length to 5 s). However, the performances of both systems degrade significantly as the duration of the utterances decreases. To address this issue, we propose two novel nonlinear mapping methods which train DNN models to map the i-vectors extracted from short utterances to their corresponding long-utterance i-vectors. The mapped i-vector can restore missing information and reduce the variance of the original short-utterance i-vectors. The proposed methods both model the joint representation of short and long utterance i-vectors: the first method trains an autoencoder first using concatenated short and long utterance i-vectors and then uses the pre-trained weights to initialize a supervised regression model from the short to long version; the second method jointly trains the supervised regression model with an autoencoder reconstructing the short utterance i-vector itself. Experimental results using the NIST SRE 2010 dataset show that both methods provide significant improvement and result in a 24.51\% relative improvement in Equal Error Rates (EERs) from a baseline system. In order to learn a better joint representation, we further investigate the effect of a deep encoder with residual blocks, and the results indicate that the residual network can further improve the EERs of a baseline system by up to 26.47\%. Moreover, in order to improve the short i-vector mapping to its long version, an additional vector, which represents the average value of phoneme posteriors across frames, is also added to the input, and results in a 28.43\% improvement. When further testing the best-validated models of SRE10 on the Speaker In The Wild (SITW) dataset, the methods result in a 23.12\% improvement on arbitrary-duration (1-5 s) short-utterance conditions.
\end{abstract}


\end{frontmatter}


\section{Introduction}


The i-vector based framework has defined the state-of-the-art for text-independent speaker recognition. The i-vectors are extracted from either a Gaussian mixture model (GMM) based \citep{Dehak:11} or a deep neural network (DNN) based system \citep{Lei:11}, and for the backend, probabilistic linear discriminant analysis (PLDA) \citep{Prince:07} has been widely used. The i-vector/ PLDA system performs well if long (e.g. more than 30 s) enrollment and test utterances are available, but the performance degrades rapidly when only limited data are available \citep{Kanagasundaram:11}. To address this issue, a range of techniques has been studied on different aspects of this problem \citep{Poddar:17,Das:17}.

There has been a number of methods to model the variation of short utterance i-vectors. In
\citet{Cumani:14,Cumani:15}, a Full Posterior Distribution PLDA (FP-PLDA) is proposed to exploit the covariance of the i-vector distribution, which improves the standard Gaussian PLDA (G-PLDA) model by accounting for the uncertainty of i-vector extraction. In \citet{Hasan:13}, the effect of short utterance i-vectors on system performance was analyzed, and the duration variability was modeled as additive noise in the i-vector space. The work in \citet{Kanagasundaram:14} introduces a short utterance variance normalization technique and a short utterance variance modeling approach at the i-vector feature level; the technique makes use of the covariance matrices of long and short i-vectors for normalization.

Alternatively, several approaches have been proposed that leverage phonetic information to perform content matching. The work in \citet{Li:16} proposes a GMM based subregion framework where speaker models are trained for each subregion defined by phonemes. Test utterances are then scored with subregion models.  In \citet{Chen:16}, the authors use the local session variability vectors estimated from certain phonetic components instead of computing the i-vector from the whole utterance. Phonetic classes are obtained by clustering similar senones (group of triphones with similar acoustic properties) that are estimated from posterior probabilities of a DNN trained for phone state classification. Another approach was proposed in \citet{Scheffer:14} which matches the zero-order statistics of test and enrollment utterances using posteriors of each phone state before computing the i-vectors.

In addition, a few studies have focused on the role of feature extraction and score calibration. In \citet{Guo:16, Guo:17a}, the authors proposed several different methods (DNN and linear regression models) to estimate speaker-specific subglottal acoustic features, which are more stationary compared to MFCCs, largely phoneme independent, and can alleviate the phoneme mismatch between training and testing utterances. In addition, \citet{Hasan:13} proposes a Quality Measure Function (QMF) which is a score-calibration mechanism that compensates for the duration mismatch in trial scores.

Recently, several approaches have been proposed which use deep neural networks to learn speaker embedding from short-utterances. In \citet{Snyder:17}, the authors use a neural network, which is trained to discriminate between a large number of speakers, to generate fixed-dimensional speaker embedding, and the speaker embedding are used for PLDA scoring. In \citet{Zhang:17}, the authors propose an end-to-end system which directly learns a speaker discriminative embedding using a triplet loss function and an Inception Net. Both methods show improvement over GMM-based i-vector systems.

A few recent papers have focused on i-vector mapping, which maps the short utterance i-vector to its long version. In \citet{Kheder:16, Kheder:18}, the authors proposed a probabilistic approach, in which a GMM-based joint model between long and short utterance i-vectors was trained, and a minimum mean square error (MMSE) estimator was applied to transform a short i-vector to its long version. Since the GMM-based mapping function is actually a weighted sum of linear functions, our previous research \citep{Guo:17b} demonstrates that a proposed non-linear mapping using convolutional neural networks (CNNs) outperforms the GMM-based linear mapping methods across different conditions. The CNN-based mapping methods use unsupervised learning to regularize the supervised regression model, and result in significant performance improvement.

This paper is an extension of our aforementioned work in \citet{Guo:17b} where we investigate neural network based non-linear mapping methods for i-vector mapping. Here, we first compare and analyze the performance of both GMM- and DNN- based i-vector systems with short-utterance evaluation tasks. Based on the results which show that I-vector\_DNN systems outperform I-vector\_GMM systems across durations, we first investigate our proposed non-linear i-vector mapping methods using I-vector\_DNN systems. Two novel DNN-based i-vector mapping methods are proposed and compared. They both model the joint representation of short and long utterance i-vectors by making use of an autoencoder.

The first method trains an autoencoder using concatenated short and long utterance i-vectors and then the pre-trained weights are used to perform fine-tuning for the supervised regression task which directly maps short to long utterances. By learning a joint embedding of short and long utterances i-vectors, the pre-trained autoencoder can help to initialize the weights at a desirable basin of the landscape of the loss function for the supervised training. Such pre-training proves to be useful especially when the training dataset is not large enough. Similar ideas of pre-training have been studied by \citet{Hinton:06} and \citet{Erhan:10}.

The second method jointly trains the supervised regression model with an autoencoder to reconstruct the short-utterance i-vector itself. The autoencoder here plays the role of a regularizer, which is important when the training dataset is not large enough and the dimensions of the input and output are relatively high. The fact that the autoencoder loss helps prevent overfitting has been observed in the machine learning literature. For example, in \citet{Rasmus:15,Zhang:16}, a supervised neural network is augmented with decoding pathways for reconstruction, and it is shown that the reconstruction loss helps improve the performance of supervised tasks. More recently, a paper on CapNet \citep{Sabour:17} introduces a decoder that plays a critical role in achieving the state of the art performance on a classification task.

We further discuss several key factors of the proposed DNN mapping models in detail, including pre-training iteration, regularization weights and encoder depth. The best model provides more than 26.47\% relative improvement. We also show that by adding additional phoneme information as input, we can achieve further mapping improvements (28.43\%). We apply the proposed mapping methods to different durations of evaluation utterances to represent real-life situations, and the results show their effectiveness across all conditions. The mapping results for both I-vector\_GMM and I-vector\_DNN systems are compared, and show significant improvement for both systems. In the end, in order to show the generalization of the proposed methods, we apply the best-validated models of SRE10 \citep{Martin:10} dataset to the Speaker In The Wild (SITW) dataset \citep{McLaren:15}, which also show considerable improvement (23.12\%).

This paper is structured as follows. Section 2 describes the state-of-the-art i-vector/PLDA speaker verification systems. Section 3 analyzes the effect of utterance duration on i-vectors and introduces the proposed DNN-based i-vector mapping methods in detail. Section 4 presents the experimental set-up. Experimental results and analysis of the proposed techniques are presented in Section 5. Section 6 discusses mapping effects, and finally, in Section 7, major conclusions are presented.

\section{I-vector based speaker verification systems}
As mentioned earlier, the state-of-the-art text-independent speaker verification system is based on the i-vector framework. In these systems, a universal background model (UBM) is used to collect sufficient statistics for i-vector extraction, and a PLDA backend is adopted to obtain the similarity scores between i-vectors. There are two different ways to model a UBM: using unsupervised-trained GMMs or using a DNN trained as a senone classifier. Therefore, we will introduce both the I-vector\_GMM and I-vector\_DNN systems as well as PLDA modeling.
\subsection{I-vector\_GMM system}
The i-vector representation is based on the total variability modeling concept which assumes that speaker- and channel- dependent variabilities reside in a low-dimensional subspace, represented by the total variability matrix $\bm{T}$. Mathematically, the speaker- and channel-dependent GMM supervector $\bm{s}$ can be modeled as:
\begin{equation}
\bm{s} = \bm{s'} + \bm{Tw}
\label{eq:1}
\end{equation}
where $\bm{s'}$ is the speaker- and channel-independent supervector, $\bm{T}$ is a rectangular matrix of low rank and $\bm{w}$ is a random vector called the i-vector which has a standard normal distribution $\mathcal{N} (0,\bm{I})$.

In order to learn the total variability subspace, the Baum-Welch statistics need to be computed for a given utterance, which are defined as:
\begin{equation}
N_c = \sum_{t}P(c|\bm{y}_t,\Omega)
\label{eq:1}
\end{equation}
\begin{equation}
\bm{F}_c = \sum_{t}P(c|\bm{y}_t,\Omega)\bm{y}_t
\label{eq:1}
\end{equation}
where $N_c$ and $\bm{F}_c$ represents the zeroth and first order statistics, $\bm{y}_t$ is the feature sample at time index $t$, $\Omega$ represent the UBM of C mixture components, $c = 1,...,C$ is the Gaussian index and $P(c|\bm{y}_t,\Omega)$ corresponds to the posterior of mixture component c generating the vector $\bm{y}_t$.

\subsection{I-vector\_DNN system}

As mentioned in the previous section, for an I-vector\_GMM system, the posterior of mixture component $c$ generating the vector $\bm{y}_t$ is computed with a GMM acoustic model trained in an unsupervised fashion (i.e. with no phonetic labels).
\begin{equation}
P(c|\bm{y}_t,\Omega) \Rightarrow P(c|\bm{y}_t,\Theta)
\label{eq:1}
\end{equation}
 However, recently, inspired by the success of DNN acoustic models in automatic speech recognition (ASR), \citet{Lei:11} proposed a method which uses DNN senone (cluster of context-dependent triphones) posteriors to replace the GMM posteriors as illustrated in Eq.4, which leads to significant improvement in speaker verification. $\Theta$ represents the trained DNN model for senone classfication.

The senone posterior approach uses ASR features to compute the class soft alignment and the standard speaker verification features for sufficient statistic estimation. Once sufficient statistics are accumulated, the training procedure is the same as in the previous section. In this paper, we use a state-of-the-art time delay neural network (TDNN) as in \citet{Peddinti:15} to train the ASR acoustic model.

\subsection{PLDA modeling}
PLDA is a generative model of i-vector distributions for speaker verification. In this paper, we use a simplified variant of PLDA, termed as G-PLDA \citep{Kenny:13}, which is widely used by researchers. A standard G-PLDA assumes that the i-vector $\bm{w}_i$ is represented by:

\begin{equation}
\bm{w}_i = \bm{r} + \bm{Ux} + \bm{\epsilon}_i
\label{eq:1}
\end{equation}
where, $\bm{r}$ is the mean of i-vectors, $\bm{U}$ defines the between-speaker subspace, and the latent variable $\bm{x}$ represents the speaker identity and is assumed to have standard normal distribution. The residual term $\bm{\epsilon}_i$ represents the within-speaker variability, which is normally distributed with zero mean and full covariance $\bm{\Sigma'}$.

PLDA based i-vector system scoring is calculated using the log likelihood ratio (LLR) between a target and test i-vectors, denoted as $\bm{w}_{target}$ and $\bm{w}_{test}$. The likelihood ratio can be calculated as follows:

\begin{equation}
LLR=\log\frac{P(\bm{w}_{target},\bm{w}_{test}|H_1)}{P(\bm{w}_{target}|H_0)P(\bm{w}_{test}|H_0)}
\label{eq:1}
\end{equation}
where $H_1$ and $H_0$ denote the hypothesis that two i-vectors represent the same speaker, and different speakers, respectively.

\section{Short-utterance speaker verification}
\begin{table*}[!t]
\centering
\setlength\tabcolsep{7pt}
\caption{Mean variance of long and short utterances (from SRE and Switchboard dataset)}
\begin{tabular} {l l l  }\hline
 & \multicolumn{2}{c}{i-vectors}
   \\\cmidrule{2-3}
   & long utterance & short utterance  \\
  \hline
 \hline
mean variance($\sigma_{mean}$) & 283   &  493  \\
 \hline
 \hline
\end{tabular}
\label{t_mv_ls_utt}
\end{table*}
\subsection{The effect of utterance durations on i-vectors}
Full-length i-vectors have relatively smaller variations compared with i-vectors extracted from short utterances \citep{Poddar:17}, because i-vectors of short utterances can vary considerably with changes in phonetic content. In order to show the variation changes between long and short utterance i-vectors, we first calculate the average diagonal covariance (denoted as $\sigma_{m}$) of i-vectors across all utterances of a given speaker $m$ and then calculate the mean (denoted as $\sigma_{mean}$) of the covariances over all speakers. $\sigma_{m}$ and $\sigma_{mean}$ are defined in Eqs.7-8 as:

\begin{equation}
\sigma_{m} = \frac{1}{N}\Sigma_{n=1}^{N}\Tr((\bm{w}_{mn}-\bar{\bm{w}_m})(\bm{w}_{mn}-\bar{\bm{w}_m})^T)
\label{eq:1}
\end{equation}

\begin{equation}
\sigma_{mean} = \frac{1}{M}\Sigma_{m=1}^{M}\sigma_m
\label{eq:1}
\end{equation}
where $\bar{\bm{w}_m}$ corresponds to the mean of the i-vectors belonging to speaker $m$. $N$ represents the total number of utterances for speaker $m$, $\Tr(.)$ represents the trace operation, and $M$ is total number of speakers.

\begin{figure}[t]
\centering
  \includegraphics[width=8cm]{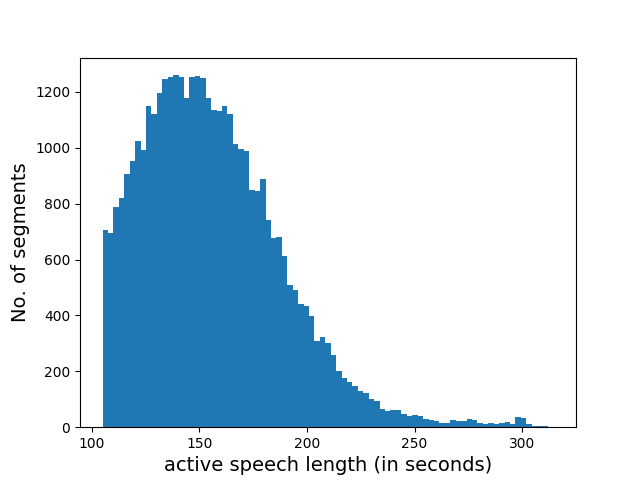}\\
  \caption{Distribution of active speech length of the selected 40000 long utterances.}\label{fig_dis_long_NIST}
\end{figure}

In order to compare the $\sigma_{mean}$ for long and short utterance i-vectors, we choose around 4000 speakers with multiple long utterances (more than 2 mins durations and 100 s active speech) from the SRE and Switchboard (SWB) datasets  (in total around 40000 long utterances) and truncate each long utterances into multiple 5-10 s short utterances. We plot the distribution of active-speech length (utterance length after voice activity detection) across these 40000 long utterances in Fig.~\ref{fig_dis_long_NIST}. The i-vectors are extracted for each short and long utterance using the I-vector\_DNN system, and Table~\ref{t_mv_ls_utt} shows the mean variance $\sigma_{mean}$ across all speakers calculated from long and short utterance i-vectors individually. The mean of variances in the Table~\ref{t_mv_ls_utt} indicates that short-utterance i-vectors have larger variation compared to those of long-utterance i-vectors.

\subsection{DNN-based i-vector mapping}

In order to alleviate possible phoneme mismatch in text-independent short utterances, we propose several methods to map short-utterance i-vectors to their long version. This mapping is a many-to-one mapping, from which we want to restore the missing information from the short-utterance i-vectors and reduce their variance.

In this section, we will introduce and compare several novel DNN-based i-vector mapping methods. Our pilot experiments indicate that, if we train a supervised DNN to learn this mapping directly, which is similar to the approaches in \citet{Bousquet:17} , the improvement is not significant, due to over-fitting to the training dataset. In order to solve this problem, we propose two different methods which both model the joint representation of short and long utterance i-vectors by using an autoencoder. The decoder reconstructs the original input representation and forces the encoded embedding to learn a hidden space which represents both short and long utterance i-vectors and thus can lead to a better generalization. The first is a two-stage method: using an autoencoder to first train a bottleneck representation of both long and short utterance i-vectors, and then uses the pre-trained weights to perform a supervised fine-tuning of the model, which maps the short-utterance i-vector to its long version directly. The second is a single-stage method: jointly train the supervised regression model with an autoencoder to reconstruct the short i-vector. The final loss to optimize is a weighted sum of the supervised regression loss and the reconstruction loss. In the following subsections, we will introduce these two methods in detail.

\subsubsection{$DNN_1$ (two-stage method): pre-training and fine-tunning}
\begin{figure}[t]
\centering
  \includegraphics[width=7.8cm]{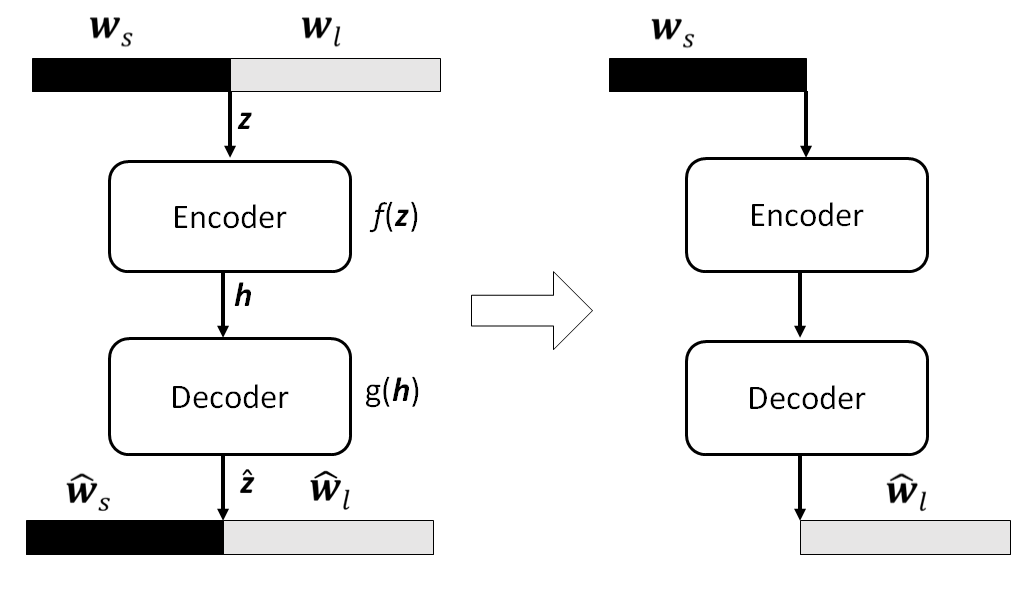}\\
  \caption{$DNN_1$: two-stage training of i-vector mapping. Left schema corresponds to the first-stage pre-training. A short-utterance i-vector  $\bm{w}_{s}$ and a corresponding long-utterance i-vector  $\bm{w}_{l}$ are first concatenated into $\bm{z}$. Then $\bm{z}$ is fed into an encoder $f(.)$ to generate the joint embedding $\bm{h}$. $\bm{h}$ is passed to the decoder $g(.)$ to generate the reconstructed $\bm{\hat{z}}$, which is expected to be a concatenation of a reconstructed $\bm{\hat{w}}_{s}$ and $\bm{\hat{w}}_{l}$. Right schema corresponds to the second-stage fine-tuning. The pre-trained weights in the first stage is used to initialize the supervised regression model from $\bm{w}_{s}$ to $\bm{w}_{l}$. After training, the estimated i-vector $\bm{\hat{w}}_{l}$ is used for evaluation.}\label{fig_two_stage}
\end{figure}

In order to find a good initialization of the supervised DNN model, we first train a joint representation of both short and long utterance i-vectors using an autoencoder. We first concatenate the short i-vector $\bm{w}_{s}$ and its long version $\bm{w}_{l}$ into $\bm{z}$, then the concatenated vector $\bm{z}$ is used to train an autoencoder with some specific constraints. The autoencoder learns the joint hidden representation of both short and long i-vectors, which leads to good initialization of the second-stage supervised fine-tuning. The autoencoder consists of an encoder and a decoder as illustrated in Fig.~\ref{fig_two_stage}. The encoder function $\bm{h} = f(\bm{z})$ learns a hidden representation of input vector $\bm{z}$, and the decoder function $\bm{\hat{z}} = g(\bm{h})$ produces a reconstruction. The learning process is described as minimizing the loss function $L(\bm{z},g(f(\bm{z})))$.
In order to learn a more useful representation, we add a restriction on the autoencoder: constrain the hidden representation $\bm{h}$ to have a relatively small dimension in order to learn the most salient features of the training data.

\begin{figure}[t]
\centering
  \includegraphics[width=5.0cm]{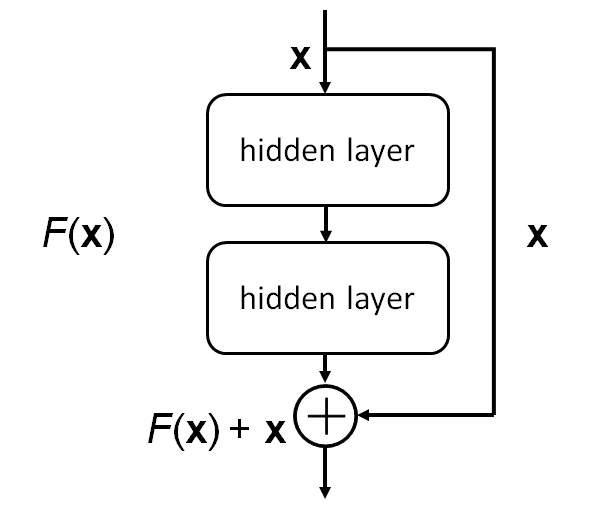}\\
  \caption{Residual block. An input $\bm{x}$ is first passed into two hidden layers to get $\bm{F(x)}$ and it also goes through a short-cut connection, which skips the hidden layers and directly comes to the output. The final output of the residual block is a summation of $\bm{F(x)}$ and $\bm{x}$.}\label{fig_res_block}
\end{figure}

\begin{figure}[!t]
\centering
  \includegraphics[width=5.0cm]{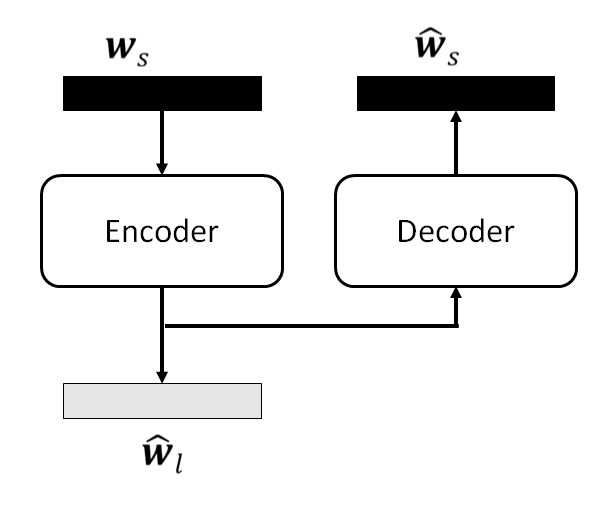}\\
  \caption{$DNN_2$: single-stage training of i-vector mapping. A short-utterance i-vector $\bm{w}_{s}$ is passed to an encoder and the output of the encoder is first used to generate the estimated long-utterance i-vector $\bm{\hat{w}}_{l}$ and it is also fed into a decoder to generate the reconstructed short-utterance i-vector $\bm{\hat{w}}_{s}$. The two tasks are optimized jointly. }\label{fig_single_stage}
\end{figure}

For the encoder function $f(.)$, we adopt options from several fully-connected layers to stacked residual blocks \citep{He:16}, in order to investigate the effect of encoder depth. Each residual block has two fully-connected layers with a short-cut connection as shown in Fig.~\ref{fig_res_block}. By using residual blocks, we are able to train a very deep neural network without adding extra parameters. A deep encoder may help learn better hidden representations. For a decoder function $g(.)$, we use a single fully connected layer with a linear regression layer, since it is enough to approximate the mapping from the learned hidden representation $\bm{h}$ to the output vector. For the loss function, we use the mean square error criterion, which is $ \| g(f(\bm{z}))-\bm{z} \|{^2}$.

Once the autoencoder is trained, we use the trained DNN-structure and weights to initialize the supervised mapping. We optimize the loss between the predicted long i-vector and the real long i-vector $\frac{1}{N} \sum_{n =1}^{N}  \| \bm{\hat{w}}_{l} - \bm{w}_{l} \|{^2}$ as shown in Fig.~\ref{fig_two_stage}. We denote this method as $DNN_1$.

\subsubsection{$DNN_2$ (single-stage method): semi-supervised training}

The two-stage method mentioned in the previous section, needs to first train a joint representation using the autoencoder and then perform a fine-tuning to train the supervised mapping. In this section, we introduce another unified semi-supervised framework based on our previous work \citep{Guo:17b} which can jointly train the supervised mapping with an autoencoder to minimize the reconstruction error. The joint framework is motivated by the fact that by sharing the hidden representations among supervised and unsupervised tasks, the network generalizes better and it can also avoid using the two-stage training procedures and speed up training. This method is denoted as $DNN_2$.

We adopt the same autoencoder framework as mentioned in the previous section, which has an encoder and a decoder, but the input to the encoder here is the short-utterance i-vector $\bm{w}_{s}$. The output from the encoder will be connected to a linear regression layer to predict the long-utterance i-vector $\bm{w}_{l}$, and it will also be used to reconstruct the short-utterance i-vector $\bm{w}_{s}$ itself by inputing it into a decoder, which gives rise to the autoencoder structure. The entire framework is shown in Fig.~\ref{fig_single_stage}.

We define a new objective function to jointly train the network. Let us use $\bm{\hat{w}}_{l}$ and $\bm{\hat{w}}_{s}$ to represent the output from the supervised regression model and autoencoder respectively. We can define the objective loss function $ L_{total}$ which combines the loss from the regression model and the autoencoder in a weighted fashion as: \newline
\begin{equation}
  L_{total} = (1-\alpha)L_{r} + \alpha L_{a}
\end{equation} \\
where $L_{r}$ is the loss of regression model defined as
\begin{equation}
  L_{r}(\bm{{w}}_{s},\bm{w}_{l};\theta_{r}) = \frac{1}{N} \sum_{n =1}^{N}  \| \bm{\hat{w}}_{l} - \bm{w}_{l} \|{^2}
\end{equation}
  and $L_{a}$ is the loss of an autoencoder defined as:
\begin{equation}
 L_{a}(\bm{{w}}_{s},\bm{w}_{s} ; \theta_{a}) = \frac{1}{N} \sum_{n =1}^{N} \|                                \bm{\hat{w}}_{s} - \bm{w}_{s} \|{^2}.
\end{equation}

Moreover, $\theta_{r}$ and $\theta_{a}$ are parameters of the regression model and autoencoder respectively, which are jointly trained and share the weights of the encoder layer. $\alpha$ is a scalar weight, which determines how much the reconstruction error is used to regularize the supervised learning. The reconstruction loss of the autoencoder $L_{a}$ forces the hidden vector generated from the encoder to reconstruct the short-utterance i-vector $\bm{w}_{s}$ in addition to predicting the target long-utterance i-vector $\bm{w}_{l}$, and helps prevent the hidden vector from over-fitting $\bm{w}_{l}$. For testing, we only use the output from the regression model $\bm{\hat{w}}_{l}$ as the mapped i-vector.

\begin{figure}[t]
\centering
  \includegraphics[width=3.7cm]{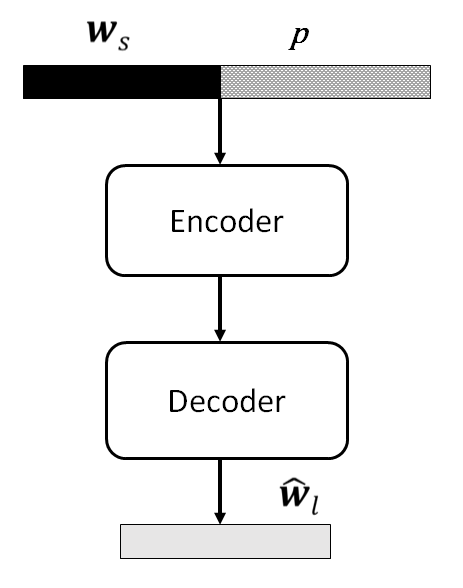}\\
  \caption{I-vector mapping with additional phoneme information. A short-utterance i-vector $\bm{w}_{s}$ is concatenated with a phoneme vector $\bm{p}$ to generate the estimated long-utterance i-vectors $\bm{\hat{w}}_{l}$.}\label{fig_phone_mapping}
\end{figure}
\begin{table*}[!t]
\centering
\setlength\tabcolsep{7pt}
\caption{Datasets used for developing I-vector\_GMM and I-vector\_DNN systems}
\begin{tabular} {l l l  }
   \hline
   \cmidrule{1-3}
   & I-vector\_GMM & I-vector\_DNN  \\
  \hline
 \hline
UBM (3472) & Switchboard, NIST 04, 05, 06, 08   &  Fisher English \\
 \hline
T (600) & Switchboard, NIST 04, 05, 06, 08   &  Switchboard, NIST 04, 05, 06, 08  \\
 \hline
PLDA & NIST 04, 05, 06, 08   &  NIST 04, 05, 06, 08  \\
 \hline
 \hline
\end{tabular}
\label{t_dev_dataset}
\end{table*}
\subsubsection{Adding phoneme information}

The variance of short utterances is mainly due to phonetic differences. In order to aid the neural network to train this non-linear mapping, for a given utterance, we extract the senone posteriors for each frame and calculate the mean posterior across frames as a phoneme vector, which is then appended to a short utterance i-vector as input (Fig.~\ref{fig_phone_mapping}). The training procedure still follows the proposed joint modeling methods ($DNN_1$ or $DNN_2$). The phoneme vectors are expected to help normalize the short-utterance i-vector, and provide extra information for this mapping. The phoneme vector $\bm{p}$ is defined as:

\begin{equation}
\bm{p} = \frac{1}{N}\sum_{t=1}^NP(c|\bm{y}_t,\Theta)
\label{eq:1}
\end{equation}

The posterior $P(c|\bm{y}_t,\Theta)$ is generated from the TDNN-based senone classifier, which was defined in Section 2.2.

\captionsetup[table]{font=normalsize,labelfont=normalsize}
\begin{table*}[th!]
\centering
\setlength\tabcolsep{10pt}
\caption{Baseline results for I-vector\_GMM and I-vector\_DNN systems under full-length and short-length utterances conditions reported in terms of EER, Relative Improvement (Rel Imp), minDCF.}
\begin{tabular} {l l l l l}\hline
 & \multicolumn{2}{c}{Female} & \multicolumn{2}{c}{Male}\\\cmidrule{2-5}
 & EER (Rel Imp)  & DCF08/DCF10 &  EER (Rel Imp)  & DCF08/DCF10  \\
 \hline
 \hline
 \multicolumn{3}{l}{Full-length condition} \\
 \hline
  I-vector\_GMM & 2.2 & 0.011/0.043 & 1.7 & 0.008/0.036  \\
  I-vector\_DNN & 1.4 (36.36\%)  & 0.005/0.022  & 0.8 (52.94\%)  & 0.003/0.017 \\
 \hline
 \hline
 \multicolumn{3}{l}{10 s-10 s condition} \\
 \hline
  I-vector\_GMM & 13.8  & 0.063/0.097 & 13.3 & 0.057/0.099 \\
  I-vector\_DNN & 12.2 (11.59\%)  & 0.054/0.093 & 10.2 (23.31\%)  & 0.048/0.095 \\
 \hline
 \hline
 \multicolumn{3}{l}{5 s-5 s condition} \\
 \hline
  I-vector\_GMM & 21.7 & 0.083/0.099  & 20.4  & 0.080/0.100\\
  I-vector\_DNN & 19.9 (8.29\%)  & 0.078/0.099 & 17.0 (16.67\%)  & 0.072/0.100 \\
 \hline
 \hline
\end{tabular}
\label{t_baseline_all}
\end{table*}

\section{Experimental set-up}

\subsection{I-vector baseline systems}
We evaluate our techniques using the state-of-the-art GMM- and DNN-based i-vector/G-PLDA systems using the Kaldi toolkit \citep{Povey:11}.

\subsubsection{Configurations of I-vector\_GMM system}
For the I-vector\_GMM system, the first 20 MFCC coefficients (discarding the zeroth coefficient) and their first and second order derivatives are extracted from the detected speech segments after an energy-based voice activity detection (VAD). A 20 ms Hamming window, a 10 ms frame shift, and a 23 channels filterbank are used. Universal background models with 3472 Gaussian components are trained, in order to have a fair comparison with the I-vector\_DNN system, whose DNN has 3472 outputs. Initial training consists of four  iterations of EM using a diagonal covariance
matrix and then an additional four iterations with a full-covariance matrix. The total variability subspace with low rank (600) is trained for five iterations of EM. The backend training consists of i-vector mean subtraction and length normalization, followed by PLDA scoring.

The UBM and i-vector extractor training data consist of male and female utterances from the SWB and NIST SRE datasets. The SWB data contains 1000 speakers and 8905 utterances of SWB 2 Phases II. The SRE dataset consists of 3805 speakers and 36614 utterances from SRE 04, 05, 06, 08. The PLDA backends are trained only on the SRE data. The dataset information is summarized in Table~\ref{t_dev_dataset}.

\subsubsection{Configurations of I-vector\_DNN system}
For the I-vector\_DNN system, a TDNN is trained using about 1,800 hours of the English portion of Fisher \citep{Cieri:04}. In the TDNN acoustic modeling system, a narrow temporal context is provided to the first layer and context width increases for the subsequent hidden layers, which enables higher levels of the network to learn greater temporal relationships. The features are 40 mel-filterbank features with a frame-length of 25 ms. Cepstral mean subtraction is performed over a window of 6 s. The TDNN has six layers, and a splicing configuration similar to those described in \citet{Peddinti:15}. In total, the DNN has a left-context of 13 and a right-context of 9. The hidden layers use the p-norm (where p = 2) activation function \citep{Zhang:14}, an input dimension of 350, and an output dimension of 3500. The softmax output layer
computes posteriors for 3472 triphone states, which is the same as the number of components for I-vector\_GMM system. No fMLLR or i-vectors are used for speaker adaptation.

The trained TDNN is used to create a UBM which directly models phonetic content.
A supervised-GMM with full-covariance is created first to initialize the i-vector extractor based on TDNN posteriors and
speaker recognition features. Training the $\bm{T}$ matrix also requires TDNN posteriors and speaker recognition features. During i-vector extraction, the only difference between this and the standard GMM-based systems is the model used to compute posteriors. In the I-vector\_GMM system, speaker recognition features are selected using a frame-level VAD, however, in order to maintain the correct temporal context, we cannot remove frames from the TDNN input features. Instead, the VAD results are used to filter out posteriors corresponding to non-speech frames.

\subsubsection{Evaluation databases}
We first evaluate our systems on condition 5 (extended task) of SRE10 \citep{Martin:10}. The test consists of conversational telephone speech in enrollment and test utterances. There are 416119 trials, over 98\% of which are nontarget comparisons. Among all trials, 236781 trials are for female speakers and 179338 trials are for male speakers. For short-utterance speaker verification tasks, we extracted short utterances which contain 10 s and 5 s speech (after VAD) from condition 5 (extended task). We train the PLDA and evaluate the trials in a gender-dependent way.

Moreover, in order to validate our proposed methods in real conditions and demonstrate the models' generalization, we use SITW, a recently published speech database  \citep{McLaren:15}. The SITW speech data was collected from open-source media channels with considerable mismatch in terms of audio conditions. We designed an arbitrary-length short-utterance task using SITW dataset to represent real-life conditions. We show the evaluation results using the best-performed models validated on SRE10 dataset.

\subsection{I-vector mapping training}
In order to train the i-vector mapping model, we selected 39754 long utterances, each having more than 100 s of speech after VAD, from the development dataset. For each long utterance, we used a 5 s or 10 s window to truncate the utterance, and the shift step is half of window size (2.5 s or 5 s). We applied the aforementioned procedures to all long utterances, and in the end we got 1.2M 10 s utterances and 2.4M 5 s utterances. All short-utterance i-vector together with its corresponding long-utterance i-vector are used as training pairs for DNN-based mapping models. We train the mapping models for each gender separately and evaluate the model in a gender-dependent way.

For the proposed two DNN-based mapping models, we use the same encoder and decoder configurations. For the encoder, we first use two fully-connected layers. The first layer has 1200 hidden nodes and the second layer has 600 hidden nodes which is a bottleneck layer (1.44M parameters in total). In order to investigate the depth of the encoder, we design a deep structure with two residual blocks and a bottleneck layer, in a total of 5 layers. Each residual block (as defined in Section 3.2.1) has two fully connected layers with 1200 hidden nodes and the bottleneck layer has 600 hidden nodes (5.76M parameters in total). For the decoder, we always use one fully-connected layer (1200 hidden nodes) with a linear output layer (1.44M parameters in total).

In order to add phoneme information for i-vector mapping, phoneme vectors are generated for each utterance by taking the average of the senone posteriors across frames. Since the phoneme vectors have a different value range compared with i-vectors, it will de-emphasize their effect for training the mapping. Therefore we scale up the phoneme vector values by a factor of 500, in order to match the range of i-vector values. The up-scaled phoneme vector is then concatenated with short-utterance i-vector for i-vector mapping.

All neural networks are trained using the Adam optimization strategy \citep{Kingma:14} with mean square error criterion and exponentially decaying learning rate starting from 0.001. The networks are initialized with the Xavier initializer \citep{Glorot:10}, which is better than the Gaussian initializer as shown in \citet{Guo:17b}. The relu activation function is used for all layers. For each layer, before passing the tensors to the nonlinearity function, a batch normalization layer \citep{Ioffe:15} is applied to normalize the tensors and speed up the convergence. For the combined loss of $DNN_2$, we set equal weights ($\alpha$=0.5) for both regression and autoencoder loss for initial experiments. The shuffling mechanism is applied on each epoch. The Tensorflow toolkit \citep{Abadi:16} is used for neural network training.
\begin{table*}[th!]
\centering
\setlength\tabcolsep{7pt}
\caption{Results for baseline (I-vector\_DNN), matched-length PLDA training, LDA dimension reduction, DNN direct mapping and proposed DNN mapping in the 10 s-10 s condition.}
\begin{tabular} {l l l l l }\hline
 & \multicolumn{2}{c}{Female} & \multicolumn{2}{c}{Male}
   \\\cmidrule{2-5}
   & EER (Rel Imp) & DCF08/DCF10  &  EER (Rel Imp) & DCF08/DCF10   \\
  \hline
 \hline
 baseline  & 12.2 & 0.054/0.093   &  10.2  & 0.048/0.095    \\
 matched length PLDA &  11.3 (7.38\%)  & 0.052/0.093   & 9.4 (7.84\%)  & 0.043/0.095  \\
 LDA 150 &  11.6 (5.00\%) & 0.052/0.093   & 9.8 (3.92\%)  & 0.047/0.093 \\
 DNN direct mapping &  10.5 (13.93\%)  & 0.054/0.096   & 9.7 (4.90\%)  & 0.047/0.093 \\
 DNN1 mapping &  9.5 (22.13\%)  & 0.047/0.091    & 7.7 (24.51\%)   & 0.039/0.090  \\
 DNN2 mapping &  9.5 (22.13\%)  & 0.047/0.091    & 7.7 (24.51\%)   & 0.039/0.089 \\
 \hline
 \hline
\end{tabular}
\label{t_comparison}
\end{table*}

\section{Evaluation results and analysis}
\subsection{I-vector baseline systems}
In this section, we present and compare two baseline systems: a I-vector\_GMM system and a I-vector\_DNN system, with standard NIST SRE 10 full-length condition and truncated 10 s-10 s and 5 s-5 s conditions.

Table~\ref{t_baseline_all} shows the equal error rate (EER) and minimum detection cost function (minDCF) of the two baseline systems under full-length evaluation condition and truncated short-length evaluation conditions. Both DCF08 and DCF10 (defined in NIST 2008 and 2010 evaluation plan) are shown in the table. From the table, we can observe that the I-vector\_DNN system gives significant improvement under the full-length condition compared with I-vector\_GMM system and achieved a max of 52.94\% relative improvement for the male condition, which is consistent with previous reported results \citep{Snyder:15}. This is mainly because the DNN model provides phonetically-aware class alignments, which can better model speakers. The good performance is also due to the strong TDNN-based senone classifier, which makes the alignments more accurate and robust. When both systems were evaluated on the truncated 10 s-10 s, 5 s-5 s evaluation conditions, the performances degrade significantly compared with the full-length condition. The main reason is that when the length of the evaluation utterance is shorter, there is significant phonetic mismatch between utterances. However, the performance of the I-vector\_DNN system still outperforms the I-vector\_GMM system by 8\%-24\%, even though the improvement is not as big as the full-length condition. From the table, we can also observe that the improvement is more significant for male speakers across all conditions. It may be the fact that phoneme classification is more accurate for male speakers, which could lead to a better phoneme-aware speaker modeling. 



\subsection{I-vector mapping results}
In this section, we show and discuss the performance of the proposed algorithms when only short utterances are available for evaluation. Since from Table~\ref{t_baseline_all} we can observe better performance using I-vector\_DNN systems, we will mainly use the I-vector\_DNN system to investigate the mapping methods. We first show the results on the 10 s-10 s condition.

Previous work \citep{Kheder:16,Guo:17b} highlights the importance of duration matching in PLDA model training. For instance when the PLDA is trained using long utterances and evaluated on short utterances, there is degradation in speaker verification performance compared to PLDA trained using matched-length short utterances. Therefore, we not only show our baseline results for the PLDA trained using the regular SRE development utterances, but also show the results for the PLDA condition using truncated matched-length short utterances.

For other baseline comparison, we first apply dimensionality reduction on i-vectors using linear discriminant analysis (LDA) and reduce the dimension of i-vectors from 600 to 150. This value has been selected according to the results of previous research \citep{Cumani:16}. LDA can maximize inter-speaker variability and minimize intra-speaker variability. We train the LDA transformation matrix using the SRE development dataset, and then, perform the dimension reduction for all development utterances and train a new PLDA model. For evaluation, all i-vectors are subjected to dimensionality reduction first and then we use the new PLDA model to get similarity scores. To compare with another short-utterance compensation technique, we evaluate the i-vector mapping methods proposed in \citet{Bousquet:17}, which use DNNs to train a direct mapping from short-utterance i-vectors to the corresponding long version. Similar to \citet{Bousquet:17}, we also add some long-utterance i-vectors as input for regularization purposes.

\begin{table*}[th!]
\centering
\setlength\tabcolsep{7pt}
\caption{DNN-based mapping results using DNNs with different depths in the 10 s-10 s condition.}
\begin{tabular} {l l l l l }\hline
 & \multicolumn{2}{c}{Female} & \multicolumn{2}{c}{Male}
   \\\cmidrule{2-5}
   & EER (Rel Imp) & DCF08/DCF10 &  EER (Rel Imp) & DCF08/DCF10 \\
  \hline
 \hline
 baseline & 12.2   & 0.054/0.093  &  10.2  & 0.048/0.095   \\
 DNN1 mapping (3 layer) &  9.5 (22.13\%)  &  0.047/0.091  & 7.7 (24.51\%)  &  0.039/0.090 \\
 DNN2 mapping (3 layer) &  9.5 (22.13\%)   & 0.047/0.091  & 7.7 (24.51\%)  &  0.039/0.089\\
 DNN1 mapping (6 layer + residual block) &  9.1 (25.41\%)   & 0.046/0.091  & 7.5 (26.47\%) & 0.038/0.089 \\
 DNN2 mapping (6 layer + residual block) &  9.3 (23.77\%)  & 0.047/0.091   & 7.6  (25.49\%) &  0.038/0.089 \\
 \hline
 \hline
\end{tabular}
\label{t_res}
\end{table*}

For our proposed DNN mapping methods, we first show the mapping results for both $DNN_1 $ and $DNN_2$ with three hidden layers. Note that for mapped i-vectors, we use the same PLDA as the baseline system to get similarity scores. We further investigate the effect of pretraining iterations for $DNN_1$, the weight $\alpha$ of the reconstruction loss for $DNN_2$ and the depth of encoder, compare the results for different durations, and investigate the effect of additional phoneme information. We also compare with mapping results for both I-vector\_GMM and I-vector\_DNN systems. In the end, we test the generalization of the trained models on the SITW dataset.

Table~\ref{t_comparison} presents the results for regular PLDA training condition (baseline), matched-length PLDA condition, LDA dimetionality reduction method, DNN-based direct mapping method, DNN-based two-stage method ($DNN_1$) and DNN-based single-stage method  ($DNN_2$, $\alpha$=0.5). We observe that matched-length PLDA training gives considerable improvement compared with non-matched PLDA training (baseline), which is consistent with previous work. When training the PLDA using short-utterance i-vectors, the system can capture the variance of short-utterance i-vectors. Using LDA to do dimentionality reduction also results in some improvement, since it reduces the variance of the i-vectors. DNN-based direct mapping gives more improvement for female speakers (13.93\%) compared with male speakers (5\%) in terms of EERs, and it may be due to the fact that more training data is available for female speakers and thus the over-fitting problem is less severe for females. In the last two rows, we show the performance of our proposed DNN-based mapping methods on short-utterance i-vectors. From the results, we can observe that they both result in significant improvements over the baseline for both the EER and minDCF metrics, and they also outperform the other short-utterance compensation methods by a large margin. $DNN_1$ and $DNN_2$ methods have comparable performance, which prove the importance of learning joint representation of both short and long utterance i-vectors. The proposed methods outperform the baseline system by 22.13\% for female speakers and improve the male speaker baseline by 24.51\%. One of the advantages using $DNN_2$ is that the unified framework avoids using the two-stage training procedure, which speeds up the training. \\


\begin{table*}[th!]
\centering
\setlength\tabcolsep{7pt}
\caption{DNN-based mapping results with additional phoneme information in the 10 s-10 s condition.}
\begin{tabular} {l l l l l }\hline
 & \multicolumn{2}{c}{Female} & \multicolumn{2}{c}{Male}
   \\\cmidrule{2-5}
   & EER (Rel Imp) & DCF08/DCF10 &  EER (Rel Imp) & DCF08/DCF10  \\
  \hline
 \hline
 baseline & 12.2   & 0.054/0.093 &  10.2 & 0.048/0.095    \\
 DNN mapping (best) &  9.1 (25.41\%)  & 0.046/0.091   & 7.5 (26.47\%)  & 0.038/0.089 \\
 DNN mapping (best) + phoneme info &  8.9 (27.05\%)  & 0.046/0.090   & 7.3 (28.43\%)  & 0.037/0.090 \\
 \hline
 \hline
\end{tabular}
\label{t_pho}
\end{table*}

\subsubsection{Effect of pre-training for $DNN_1$ }
In this section, we will show how first-stage pre-training influences the second-stage mapping training for $DNN_1$. We have investigated the number of training iterations used for first-stage pre-training from 10000-50000. What we find interesting is that when  the number of training iterations is small, the second stage fine-tuning will over-fit the data, but when the number of training iterations is large, the fine-tuning results are not optimal. In the end, 25000 iterations was a roughly good initialization for second stage fine-tuning. This indicates that the number of iterations for unsupervised training does influence the second-stage supervised training. 

\begin{figure}[th]
\centering
  \includegraphics[width=7.1cm]{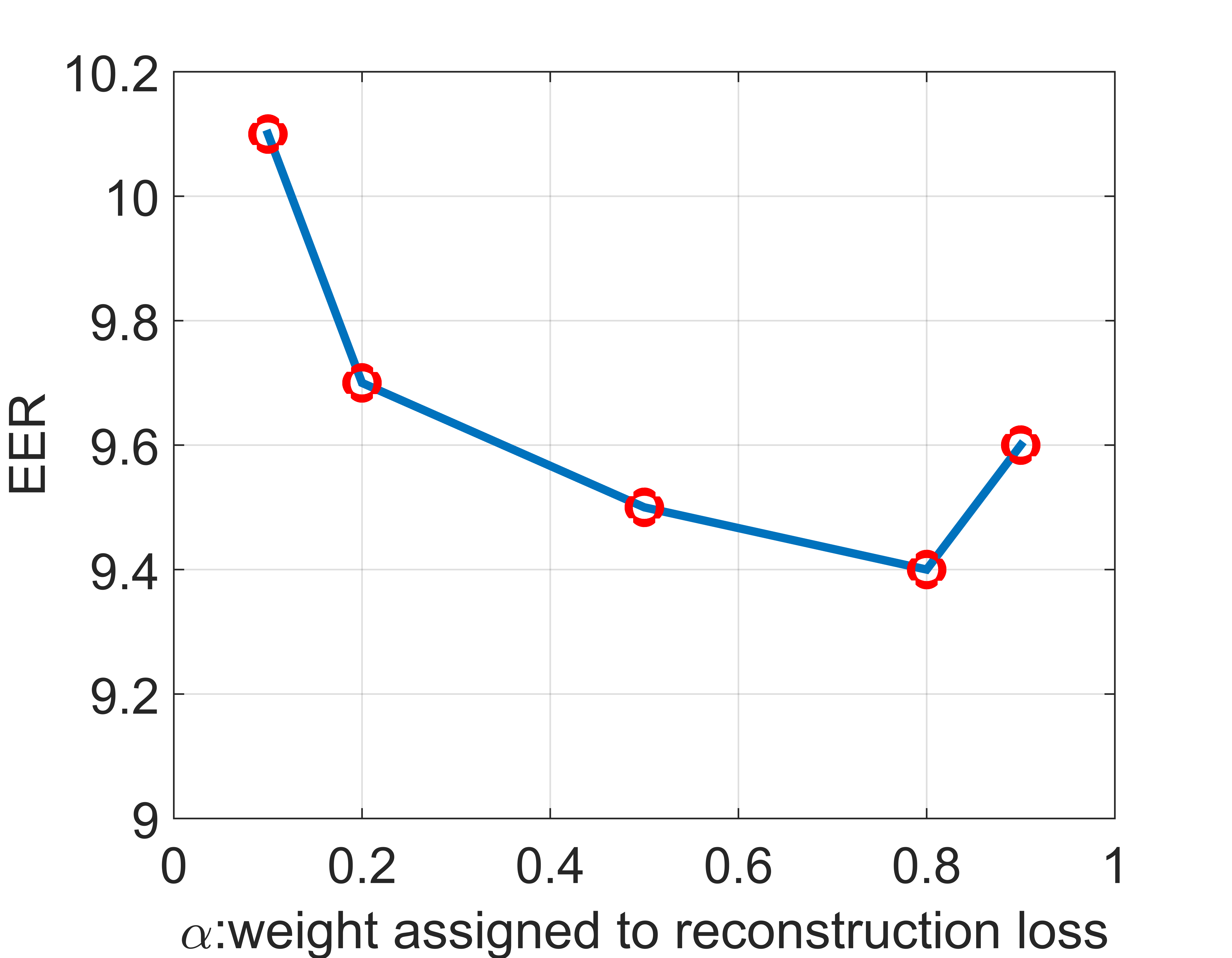}\\
  \caption{EER as a function of reconstruction loss $\alpha$ for $DNN_2$.}\label{fig_eer_weights}
\end{figure}

\subsubsection{Effect of reconstruction loss for $DNN_2$ }
In this section, we investigate the impact of the weights for the reconstruction loss in $DNN_2$. We set $\alpha$ = \{0.1,0.2,0.5,\\0.8,0.9\}. Since the weight of regression loss is $1-\alpha$, the larger $\alpha$ is, the less weight will be assigned to regression loss. Fig.~\ref{fig_eer_weights} shows the EER for female speakers as a function of the weights assigned to reconstruction loss. The reconstruction loss is clearly important for this joint learning framework. It forces the network to learn the original representations for short utterances, which can regularize the regression task and generalize better. The optimal reconstruction weight is $\alpha$ = 0.8, which indicates that the reconstruction loss is even more important for this task. Hence, it appears that unsupervised learning is very crucial for a speaker recognition task.


\begin{table*}[th!]
\centering
\setlength\tabcolsep{7pt}
\caption{DNN-based mapping results with different utterance durations.}
\begin{tabular} {l l l l l}\hline
 & \multicolumn{2}{c}{Female} & \multicolumn{2}{c}{Male}\\\cmidrule{2-5}
 & EER (Rel Imp)& DCF08/DCF10 &  EER (Rel Imp) & DCF08/DCF10  \\
 \hline
 \hline
 \multicolumn{3}{l}{10 s-10 s} \\
 \hline
  baseline & 12.2 & 0.054/0.093  & 10.2 & 0.048/0.095 \\
  DNN mapping (best) & 9.1 (25.41\%)  & 0.046/0.091  & 7.5 (26.47\%)  & 0.038/0.089 \\
 \hline
 \hline
 \multicolumn{3}{l}{5 s-5 s} \\
 \hline
  baseline & 19.9  & 0.078/0.099  & 17.0 & 0.072/0.100 \\
  DNN mapping (best) & 14.8 (25.62\%) & 0.067/0.099   & 13.5 (20.59\%)   &  0.061/0.100 \\

 \hline
 \hline
 \multicolumn{3}{l}{mix} \\
 \hline
  baseline & 17.8 & 0.068/0.097  & 14.4 & 0.061/0.100 \\
  DNN mapping (best) & 13.2 (25.84\%)  & 0.061/0.097 & 11.8 (18.06\%) & 0.053/0.096\\

    \hline
  \hline
\end{tabular}
\label{t_diff_duration}
\end{table*}

\subsubsection{Effect of encoder depth }
The depth of neural network has been proven to be important for network performance. Adding more layers will make the network more efficient and powerful to model data. Therefore, as discussed in Section 4.2, we will compare a shallow (2-layer) and a deep (5-layer) encoder for both $DNN_1$ and $DNN_2$. It's well known that training a deep model suffers a lot from gradient vanishing/exploding problems and also it can be easily stuck into local minimum points. Therefore, we use two methods to alleviate this problem. Firstly, as stated in Section 4.2, we use a normalized initialization (Xavier initialization) and a batch normalization layer to normalize the intermediate hidden output. Secondly, we apply residual learning, which uses several residual blocks (defined in Section 3.2.1) with no extra parameter compared with regular fully-connected layers. The residual blocks will make the information flow between layers easy and enable very smooth forward/backward propagation, which makes it feasible to train deep networks. To our knowledge, this is one of the first studies to investigate the effect of residual networks for auto-encoder and unsupervised learning. Here, for the deep encoder, we use 2 residual blocks and 1 fully-connected bottleneck layer (in total 5 layers). For the decoder, we use a single hidden layer with a linear regression output layer.

From Table~\ref{t_res}, we can observe that a deep encoder does result in improvements compared with a shallow encoder. Especially for $DNN_1$, the residual networks give a 25.41\% relative improvement for female speakers and 26.47\% relative improvement for male speakers. The results indicate that learning a good joint representation of both short and long utterance i-vectors is very beneficial for this supervised mapping task, and the deep encoder can help learn a better bottleneck joint embedding. The deep encoder can also decrease the amount of training data needed to model the non-linear function, which can also alleviate the over-fitting problem. In order to show the effect of residual short-cuts, we performed experiments using a deep encoder without short-cut connections, and the system resulted in even worse performance compared with the shallow encoder. Therefore, residual blocks with short-cut connections are very crucial for deep neural network training, since it alleviates the hard optimization problems of deep networks.


\subsubsection{Effect of adding phoneme information}

In this section, we show the results when adding phoneme vector (mean of phoneme posteriors across frames) with short-utterance i-vectors to learn the mapping. We will investigate the effect of adding phoneme information based on the best performed DNN-mapping structures. From Table~\ref{t_pho}, we can observe that when adding phoneme vector, the EER further improves to 8.9\% for female speakers and 7.3\% for male speakers from the previous best DNN-mapping results. It achieves the best results for this task. The results prove the hypothesis that adding a phoneme vector can help the neural network reduce the variance of short-utterance i-vectors, which will lead to better and more generalizable mapping results. In Section 5.4, we will also show the effect of adding phoneme vectors to GMM-i-vectors.

\subsection{Results with different durations}

In this section, the results for different durations of evaluation utterances are listed. Table~\ref{t_diff_duration} shows the baseline and the best mapping results for 10 s-10 s, 5 s-5 s and mixed duration conditions.
From the table, we can observe that the proposed methods give significant improvements for both 10 s-10 s and 5 s-5 s conditions, which indicates that the proposed method generalizes to different durations. In real applications, however, the duration of short utterances can not be controlled, therefore we train the mapping using the i-vectors generated from mixed 10 s and 5 s utterances and show the results also on a mixed-duration evaluation task (mixed of 5 s and 10 s). From Table~\ref{t_diff_duration}, we can see that the baseline results for the mixed condition range between the EER results of 10 s-10 s and the 5 s-5 s evaluation tasks. The proposed mapping algorithms can model i-vectors extracted from various durations, and thus give consistent improvement as shown in the table. 





\begin{table*}[th!]
\centering
\setlength\tabcolsep{7pt}
\caption{Results for  I-vector\_GMM and  I-vector\_DNN systems in the 10 s-10 s conditions.}
\begin{tabular} {l l l l l}\hline
 & \multicolumn{2}{c}{Female} & \multicolumn{2}{c}{Male}\\\cmidrule{2-5}
 & EER (Rel Imp) & DCF08/DCF10 &  EER (Rel Imp) & DCF08/DCF10  \\
 \hline
 \hline
 \multicolumn{3}{l}{I-vector\_GMM} \\
 \hline
  baseline & 13.8 & 0.063/0.097  & 13.3 & 0.057/0.099 \\
  DNN mapping (best) & 11.0 (20.29\%)  & 0.054/0.095 & 10.6 (20.30\%)  & 0.051/0.096  \\
        DNN mapping (best) + phoneme info & 10.4 (24.64\%)  & 0.053/0.094  & 9.6 (27.82\%)  & 0.048/0.096 \\
 \hline
 \hline
 \multicolumn{3}{l}{I-vector\_DNN} \\
 \hline
  baseline & 12.2  & 0.054/0.093 & 10.2 & 0.048/0.095 \\
  DNN mapping (best) & 9.1 (25.41\%) & 0.046/0.091  & 7.5 (26.47\%) & 0.038/0.089 \\
  DNN mapping (best) + phoneme info & 8.9 (27.05\%)  & 0.046/0.090 & 7.3 (28.43\%)  & 0.037/0.090 \\

    \hline
  \hline
\end{tabular}
\label{t_GMM_DNN_comp}
\end{table*}

\begin{figure*}[!h]
  \centering
  \subfigure[female speakers]{\includegraphics[scale=0.421]{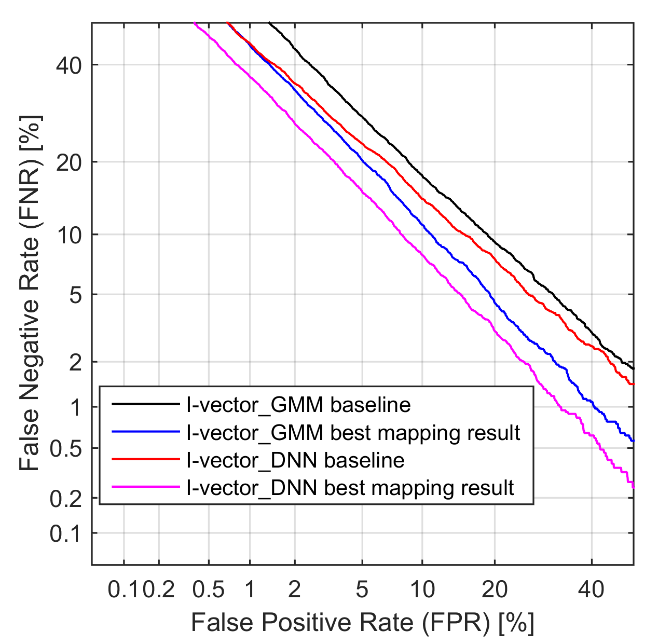}}\quad
  \subfigure[male speakers]{\includegraphics[scale=0.320]{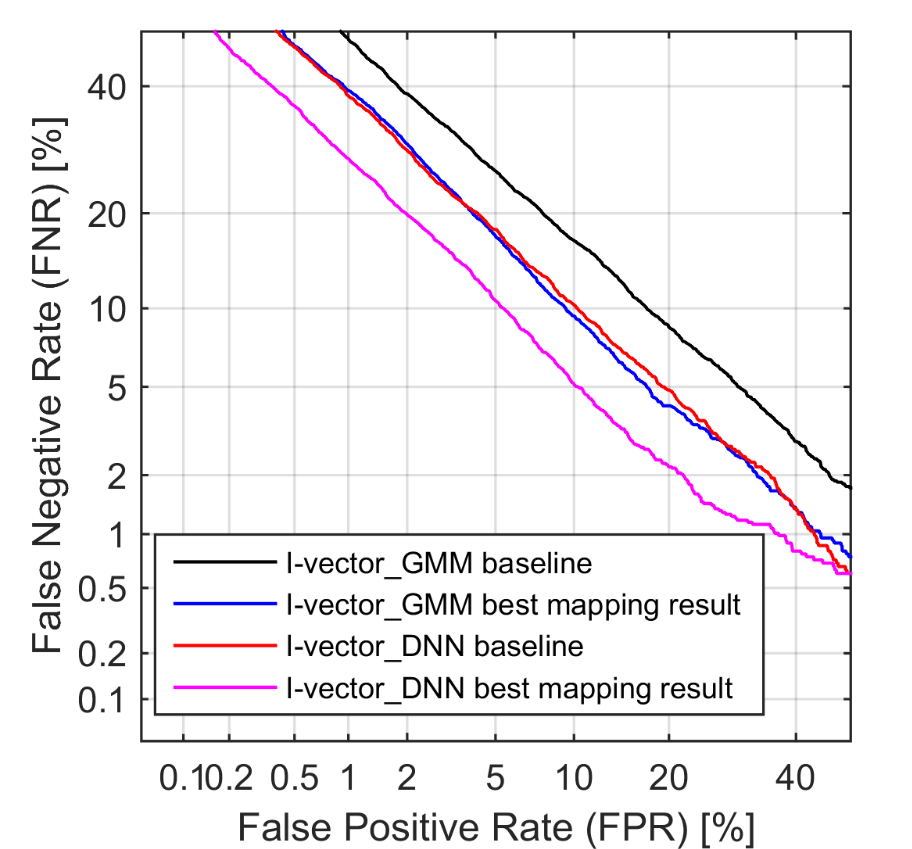}}
  \caption{DET curves for the mapping results of I-vector\_GMM and I-vector\_DNN systems under 10 s-10 s conditions. Left figure corresponds to female speakers and right one corresponds to male speakers.}\label{fig_det}
\end{figure*}

\subsection{Comparison of mapping results for both  I-vector\_GMM and I-vector\_DNN systems}
In the previous sections, we only show the mapping experiments for I-vector\_DNN system, therefore, in this section, we will show the mapping results for the I-vector\_GMM system. In Section 5.1, we show that for the baseline results, I-vector\_DNN system outperforms the I-vector\_GMM system, but it is also interesting to compare the results after mapping. From Table~\ref{t_GMM_DNN_comp} we observe that the proposed mapping methods give significant improvement for both systems. After mapping, the I-vector\_DNN systems still outperform the I-vector\_GMM systems and the superiority of I-vector\_DNN systems is even more significant. We also compare the mapping results when adding phoneme vectors. The table shows that the effect of adding phoneme information is more significant for GMM-i-vectors and it can achieve as much as a 10\% relative improvement on the best DNN mapping baseline. The reason is that DNN-i-vectors already contain some phoneme information, while GMM-i-vectors do not have clear phoneme representation. Therefore GMM-i-vectors can benefit more from adding phoneme vectors. In the end, we summarize the baseline and the best mapping results for both systems in Fig.~\ref{fig_det}. The DET (Detection Error Trade-off) curves are presented for both female and male speakers. The figures indicate that the proposed mapping algorithms give significant improvement from the baseline across all operation points.


\begin{table*}[th!]
\centering
\setlength\tabcolsep{7pt}
\caption{DNN-based mapping results on SITW using arbitrary durations of short utterances.}
\begin{tabular} {l l l l l}\hline
 & \multicolumn{2}{c}{Female} & \multicolumn{2}{c}{Male}\\\cmidrule{2-5}
 & EER (Rel Imp) & DCF08/DCF10  &  EER (Rel Imp) & DCF08/DCF10   \\
 \hline
 \hline
 \multicolumn{3}{l}{Arbitrary durations} \\
 \hline
  baseline & 17.3   & 0.061/0.089  & 12.0 & 0.046/0.083 \\
  DNN mapping (best models from SRE10) & 13.3 (23.12\%)  & 0.050/0.086 &  9.4 (21.67\%) &0.039/0.078\\

    \hline
  \hline
\end{tabular}
\label{t_SITW}
\end{table*}

\subsection{Performance on the SITW database}

\begin{figure}[t!]
\centering
  \includegraphics[width=8cm]{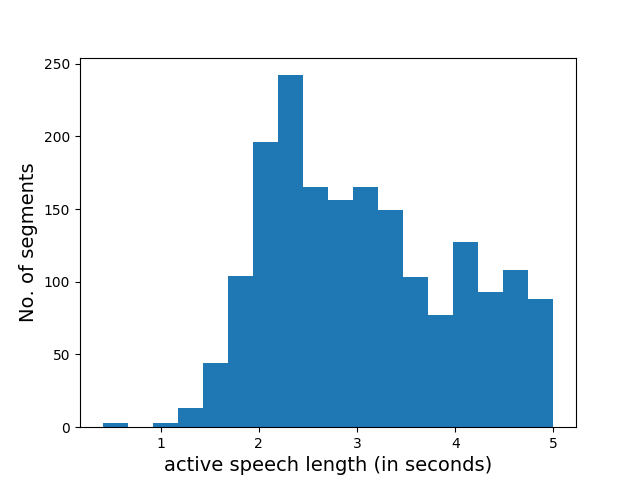}\\
  \caption{Distribution of active speech length of truncated short utterances in the SITW database.}\label{fig_dis_SITW}
\end{figure}

In the previous experiments, we show the performance of our proposed DNN-mapping methods on NIST data. In this subsection, we apply our technique on the recently published database SITW which contains real-world audio files collected from open-source media channels with considerable mismatch conditions. In order to generate a large number of random-duration short utterances, we first combined the dev and eval datasets and then selected 207 utterances from relatively clean condition. We truncated each of 207 utterances into several non-overlapped short utterances with duration 5 s, 3.5 s, 2.5 s (including both speech and non-speech portions). In the end, a total number of 1836 utterances was generated. We plot the distribution of active speech length across these 1836 utterances in Fig.~\ref{fig_dis_SITW}. From the figure, we can observe that active speech length varies between 1 s-5 s across those short utterances. Therefore, we can use these short utterances to design trials, which represent real-world conditions (arbitrary-length short utterances). In total, we designed 664672 trials for our arbitrary-length short-utterance speaker verification task. 

For each short utterance, we first down-sampled the audio files to 8 kHz sampling rate, and then extracted the i-vectors using the previously trained I-vector\_DNN system introduced in Section 4.1. For PLDA scoring, we use the same PLDA in Section 4.1, which is trained using the SRE dataset. For i-vector mapping, we use the best-validated models on SRE10 dataset (5 s condition) to apply to the SITW dataset. Evaluation results of EERs and minDCFs are show in Table~\ref{t_SITW}. From the table, we can observe that the best models validated on SRE10 dataset generalize well to the SITW dataset, which give a 23.12\% relative improvement of EERs for female speakers and a 21.67\% relative improvement for male speakers. The results also indicate that the proposed methods can be used in real-life conditions, such as smart home and forensic related applications.


\section{Mapping effects}
In order to investigate the effect of the proposed i-vector mapping algorithms, we first calculate the average square Euclidean distance between short and long utterance i-vector pairs on the SRE10 evaluation dataset before and after mapping. The average mean square Euclidean distance $D_{sl}$ between short and long utterance i-vector is defined as follow:

\begin{equation}
D_{sl} = \frac{1}{N}\Sigma_{s=1}^{N}(\Sigma_{i=1}^{L}(\bm{w}_s(i)-\bm{w}_l(i))^2)
\end{equation}
where $\bm{w}_{s}$ and $\bm{w}_{l}$ represent the short-utterance and long-utterance i-vector respectively, $L$ is the length of i-vectors and $N$ is number of short and long i-vector pairs.

We compare the $D_{sl}$ values for 10 s and 5 s short-utterance i-vectors and also the mapped 10 s and 5 s short-utterance i-vectors for female and male speakers in Table~\ref{t_e_distance}. From the table, we observe that, after mapping, the mapped short-utterance i-vectors have considerably smaller $D_{sl}$ compared to the ones before mapping. After mapping, the $D_{sl}$ in the 10 s condition is smaller compared with the 5 s condition. 

Moreover, we calculate and compare the J-ratio \citep{Fukunaga:90} of the short-utterance i-vectors from SRE10 before and after mapping in Table~\ref{t_j-ratio}, which measures the ability of speaker separation. Given i-vectors for $M$ speakers, the J-ratio can be computed using Eqs.14-16:

\begin{equation}
\bm{S}_w = \frac{1}{M}\Sigma_{s=1}^{M}\bm{R}_i
\end{equation}

\begin{equation}
\bm{S}_b = \frac{1}{M}\Sigma_{s=1}^{M}(\bm{w}_i-\bm{w}_o)(\bm{w}_i-\bm{w}_o)^T
\end{equation}

\begin{equation}
J = \Tr((\bm{S}_b+\bm{S}_w)^{-1}\bm{S}_b)
\end{equation}

where $\bm{S}_w$ is the within-class scatter matrix, $\bm{S}_b$ is the between-class scatter matrix, $\bm{w}_i$ is the mean i-vector for the $i_{th}$ speaker, $\bm{w}_o$ is the mean of all $\bm{w}_i$s, and $\bm{R}_i$ is the covariance matrix for the $i_{th}$ speaker (note that a higher J-Ratio means better separation).

From Table~\ref{t_j-ratio}, we can observe that the mapped i-vectors have considerably higher J-ratios compared with original short-utterance i-vectors for both 5 s and 10 s conditions.

These results indicate that the proposed DNN-based mapping methods can generalize well to unseen speakers and utterances, and improve the speaker separation ability of i-vectors.


\begin{table*}[!th]
\centering
\setlength\tabcolsep{7pt}
\caption{Square Euclidean distance ($D_{sl}$) between short and long utterance i-vector pairs from SRE10 before and after mapping.}
\begin{tabular} {l l l l l }\hline
& \multicolumn{4}{c}{$D_{sl}$}\\ \cmidrule{2-5}
 & \multicolumn{2}{c}{10 s} & \multicolumn{2}{c}{5 s}
   \\\cmidrule{2-5}
   & original & mapped &  original & mapped  \\
  \hline
 \hline
female & 558.3   &  306.8 &  618.8  & 352.1  \\
 \hline
 male & 493.2   &  308.8 &  556.1  & 346.5  \\
 \hline
 \hline
\end{tabular}
\label{t_e_distance}
\end{table*}

\begin{table*}[!th]
\centering
\setlength\tabcolsep{7pt}
\caption{J-ratio for short-utterance i-vectors from SRE10 before and after mapping.}
\begin{tabular} {l l l l l }\hline
& \multicolumn{4}{c}{J-ratio}\\ \cmidrule{2-5}
 & \multicolumn{2}{c}{10 s} & \multicolumn{2}{c}{5 s}
   \\\cmidrule{2-5}
   & original & mapped &  original & mapped  \\
  \hline
 \hline
female  & 87.96   &  92.97 &  82.73  & 85.18  \\
 \hline
 male & 85.23 & 90.25  & 80.41  & 84.39  \\
 \hline
 \hline
\end{tabular}
\label{t_j-ratio}
\end{table*}

\section{Conclusions}

In this paper, we show how the performance of both GMM and DNN-based i-vector speaker verification systems degrade rapidly as the duration of the evaluation utterances decreases. This paper explains and analyzes the reasons of the degradation and proposes several DNN-based techniques to train a non-linear mapping from short-utterance i-vectors to their long version, in order to improve the short-utterance evaluation performance.

Two DNN-based mapping methods ($DNN_1$ and $DNN_2$) are proposed and they both model the joint representations of short-utterance and long-utterance i-vectors. For $DNN_1$, an auto-encoder is trained first using concatenated short- and long- utterance i-vectors in order to learn a joint hidden representation, and then the pre-trained DNN is fine tuned by a supervised mapping from short to long i-vectors. $DNN_2$ adopts a unified structure, which jointly trains the supervised regression task with an auto-encoder since auto-encoders can directly regularize the non-linear mapping between short and long utterances. The unified structure simplifies the training procedure and can also learn a generalized non-linear function.

Both $DNN_1$ and $DNN_2$ result in significant improvement over the short-utterance evaluation baseline for both male and female speakers, and they also outperform other short-utterance compensation techniques by a large margin. After performing a t-test (p\textless0.001), the results indicate that all the improvements are statistically significant. We study several key factors of DNN models and conclude the following: 1) for the two-stage trained DNN model ($DNN_1$), the number of iterations for unsupervised training in the first stage is important for second-stage supervised training; 2) for the semi-supervised trained DNN model ($DNN_2$), unsupervised training plays a more important role than supervised training in a speaker verification task; 3) by increasing the depth of the neural networks using residual blocks, we can alleviate the hard optimization problem of deep neural networks and obtain an improvement compared with a shallow network, especially for $DNN_1$; 4) adding phoneme information can aid in learning the non-linear mapping and provide further performance improvement, and the effect is more significant for GMM i-vectors; 5) the proposed DNN-based mapping methods work well for short utterances with different and mixed durations; 6) the proposed models can also improve both I-vector\_GMM and I-vector\_DNN systems and after mapping, a I-vector\_DNN system still performs better than a I-vector\_GMM system; and 7) the best-validated models of SRE10 generalize well to the SITW dataset and give significant improvement for arbitrary-length short utterances. 



\end{document}